\documentclass[twocolumn,prl,aps,showpacs]{revtex4}
\usepackage{graphicx}

\begin{document}

\title{\bf Scaling Theory of Polyelectrolyte Adsorption on Repulsive Charged
Surface}

\author{Chi-Ho Cheng}
\email{phcch@phy.ncu.edu.tw}
\author{Pik-Yin Lai}

\affiliation{Department of Physics and Center for Complex Systems,
National Central University, Taiwan}

\date{\today}

\begin{abstract}
We studied the problem of many-polyelectrolyte adsorption on a
repulsive charged surface by scaling analysis. According to ratio
of the dielectric constant between the medium and the substrate,
the phase diagrams of the adsorbed layer are divided into two
classes. Their phase diagrams are qualitatively different. The
polyelectrolytes of low (high) $fZ^2$ are adsorbed for low (high)
dielectric ratio, where $f$ is the fraction of charged monomers
and $Z$ is the polyelectrolyte valency.
\end{abstract}


\maketitle

%\vspace{-5pt}

Polyelectroyte adsorption on an attractive surface has been well
studied both theoretically and experimentally for a long time
\cite{netz}. Recently the marcoion adsorption and the associated
charge inversion acquires lots of attention \cite{grosberg}.
Theoretical interests to the polyelectrolyte adsorption on an
attractive surface is mainly due to its importance for
understanding the multilayer formation by alternate deposition of
positively and negatively charged polyelectrolytes on charged
surface \cite{decher}.

Basically the theoretical works dealing with many-chain
polyelectrolyte adsorption on an attractive charged surface
consists of two different approaches. One is the self-consistent
field method, that is, to solve both Edwards equation and
Poisson-Boltzmann equation simultaneously \cite{pb1,pb2,pb3}. The
Edwards equation describes the polyelectrolyte conformation
fluctuation at the ground-state dominance limit \cite{doi} and the
Poisson-Boltzmann takes care of the local electrostatic potential
at mean-field level \cite{israelachvili}. The other is scaling
analysis \cite{deGennes}. It was predicted that the
polyelectrolyte adsorbed on an attractive surface undergoes a
transition from a compressed state to a undeformed state before
the de-sorption \cite{dobrynin}. The scaling behavior of the
adsorption layer thickness with respect to the surface charge
density is different from that obtained by the self-consistent
field method in which the overall conformational changes is
ignored.

Recently the possibility of a single chain adsorbed onto a
high-dielectric substrate with repulsive surface charge was raised
\cite{cheng}. It shows that the polyelectrolyte under the Coloumb
image attraction can overcome the slightly repulsive surface and
is able to be adsorbed on the substrate. In this paper, we would
like to investigate many-chain adsorption by scaling analysis.

%In the following, we extend the scaling approach developed by
%Dobrynin {\it et.al.} \cite{dobrynin} on the case of an attractive
%surface to the problem of a repulsive surface.

Consider a polyelectrolyte chain with degree of polymerization
$N$, fraction of charged monomers $f$, and bond length $a$ in a
solvent of dielectric constant $\epsilon$. Below the solvent,
there is a substrate of high dielectric constant
$\epsilon'>\epsilon$. The surface charge density on the substrate
is $\sigma$. At low ionic strength in which the bulk Debye
screening length is much larger than the adsorption layer
thickness, a single polyelectrolyte is still adsorbed when the
charged substrate is slightly repulsive; that is, when the surface
charge density $\sigma$ does not exceed some threshold value
$\sigma_{\rm t}$. The polyelectrolyte is going to de-sorb when the
binding energy due to image charge attraction meets the potential
barrier formed by repulsive surface \cite{cheng}.

When polyelectrolytes of multivalency $Z = f N \gg 1$ are adsorbed
onto the repulsive surface, they forms a 2d Wigner liquid on the
surface \cite{rouzina,perel,shklovskii,dobrynin2,nguyen}.
Poisson-Boltzmann theory fails because of strong correlation
between polyelectrolytes. Dobrynin {\it et.al.} developed the
scaling theory to polyelectrolyte adsorption on an attractive
surface \cite{dobrynin}. In the following, we extend the scaling
approach to the problem of a repulsive charged surface to take
care of the strong correlation between the adsorbed
polyelectrolytes. The monovalent counterions are then treated
within the Poisson-Boltzmann theory since it is generally believed
that mean-field theory is still valid for monovalent ions
\cite{grosberg}.

Suppose the surface Debye screening length, $r_{\rm s, 0}$, is
larger than the average distance between polyelectrolytes, $R$;
otherwise, any two chains cannot feel each other and the problem
is reduced to a single-chain adsorption \cite{cheng}. Notice that
the effective Debye screening length $r_{\rm s, 0}$ is not
necessary equal to the bulk one $r_{\rm s, \infty}$ because of
non-uniform counterion distribution at different height from the
charged surface.

In the presence of high dielectric substrate, the polyelectrolyte
would feel a binding energy due to image charge attraction. Its
layer thickness $D$ can be determined by the balance between
electrostatic attractive energy and conformational entropy,
\begin{equation} \label{balance}
\frac{\Gamma (Z e)^2}{\epsilon D} \simeq \frac{k_{\rm
B}Ta^2N}{D^2}
\end{equation}
where $\Gamma=(\epsilon'-\epsilon)/(\epsilon'+\epsilon)$ measures
the coupling strength between the polyelectrolyte and its image.
Eq.(\ref{balance}) gives
\begin{equation}
D \simeq a (l_{\rm B}/a)^{-1}\Gamma^{-1}f^{-1}Z^{-1}
\end{equation}
where $l_{\rm B}$ is the Bjerrum length. The smaller fraction of
charged monomers $f$, the larger conformational entropy and
adsorbed layer thickness. The binding energy is thus
\begin{equation}
W_{\rm bind} \simeq -k_{\rm B}T (l_{\rm B}/a)^2\Gamma^2 f Z^3/R^2
\end{equation}

Because of the repulsive charged surface, we have repulsive energy
from the surface
\begin{eqnarray}
W_{\rm surf} &\simeq& k_{\rm B}T \frac{(1-\Gamma)Zl_{\rm B}
(\sigma/e)}{R^2}
\int_0^\infty dr \exp(-\frac{r}{r_{\rm s, 0}}) \nonumber \\
&=& k_{\rm B}T  \frac{(1-\Gamma)Z l_{\rm B}r_{\rm s, 0}
(\sigma/e)}{R^2}
\end{eqnarray}

Repulsive energy from other polyelectrolytes
\begin{eqnarray}
W_{\rm rep}&\simeq& k_{\rm B}T \frac{(1-\Gamma)^2Z^2l_{\rm
B}}{R^4} \int_R^\infty dr \exp(-\frac{r}{r_{\rm s, 0}}) \nonumber
\\
&=& k_{\rm B}T \frac{(1-\Gamma)^2Z^2l_{\rm B}r_{ \rm D, 0
}}{R^4}\exp(-\frac{R}{r_{\rm s, 0}})
\end{eqnarray}

The average distance between polyelectrolytes is determined by
minimization of the total energy $W = W_{\rm bind}+W_{\rm
surf}+W_{\rm rep}$ with respect to $R$, which gives
\begin{eqnarray} \label{minimize}
&&\frac{\Gamma^2 f Z^2 l_{\rm B}}{a^2}-(1-\Gamma)r_{\rm s,
0}(\sigma/e) \nonumber \\ &=& \frac{2(1-\Gamma)^2 Z r_{\rm s,
0}}{R^2}(1+\frac{R}{4 r_{\rm s, 0 }})\exp(-\frac{R}{r_{\rm s, 0}})
\end{eqnarray}

In order to have a solution from Eq.(\ref{minimize}) for $\sigma
> 0$, one should have
\begin{equation}
\frac{\Gamma^2 f Z^2 l_{\rm B}}{a^2}> (1-\Gamma)r_{\rm s,
0}(\sigma/e)
\end{equation}
which determines the critical surface charge density
\begin{equation} \label{sigmac}
\sigma_{\rm c} \simeq e \frac{\Gamma^2 f Z^2l_{\rm
B}}{(1-\Gamma)a^2 r_{\rm s, 0}}
\end{equation}
$\sigma_{\rm c}$ decreases with increasing $r_{\rm s, 0}$. It is
due to more repulsion from its neighbors and surface charge when
the Debye screening length increases.

%%%%%%%%%%%%%%%%%%%%%%%%%%%%%%%%%%%%%%%%%%%%%%%%%%%%%%%%%%%%%%%%%%%%%%%%%%%%%%%%%%
%%%%%%%%%%%%%%%%%%%%%%%%%%%%%%%%%%%%%%%%%%%%%%%%%%%%%%%%%%%%%%%%%%%%%%%%%

\vspace{20pt}
\begin{figure}[tbh]
\begin{center}
\includegraphics[width=3in]{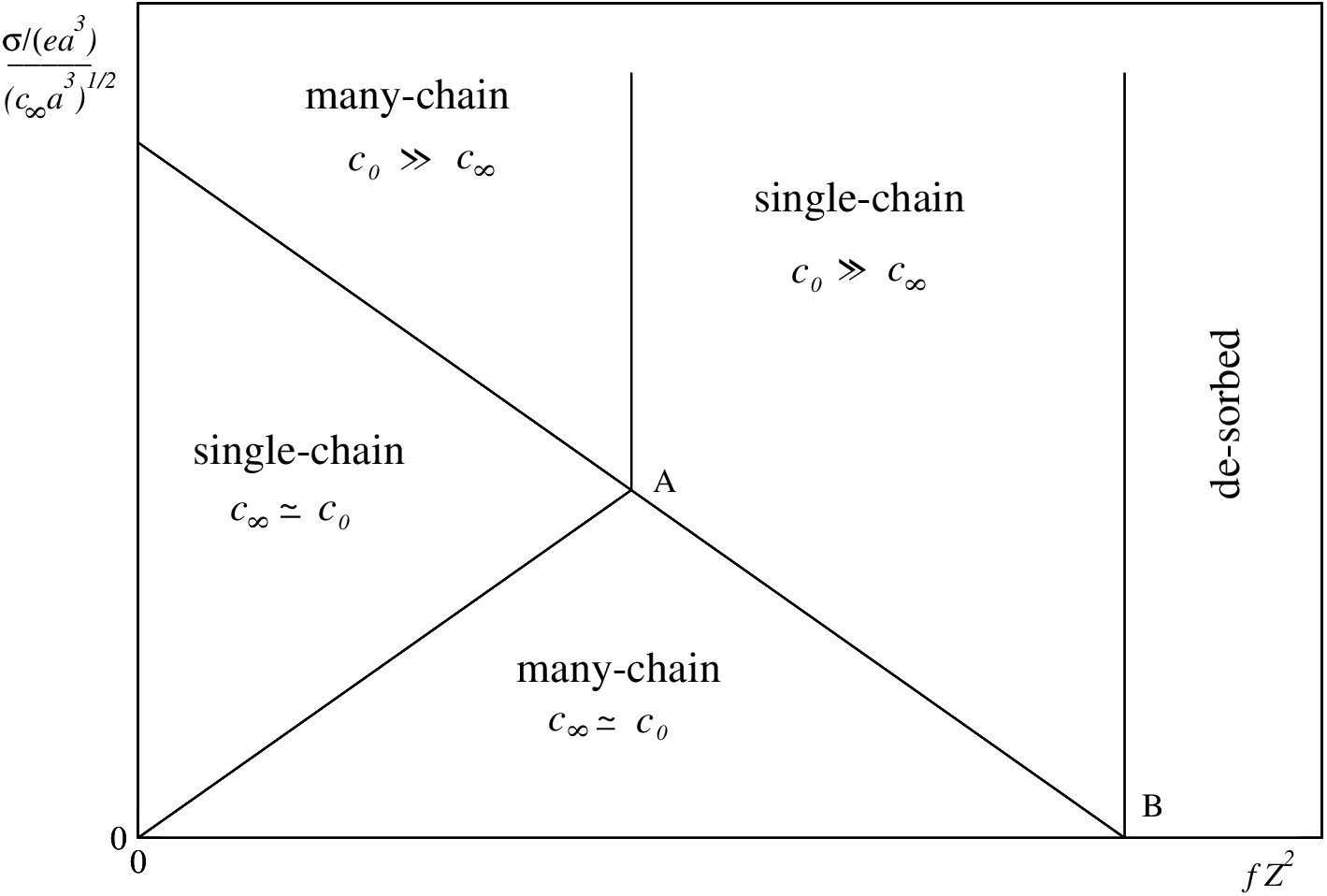}
\end{center}
\vspace{-5pt} \caption{Phase diagram of the adsorbed layer for
$\Gamma<1/2$ (or $\epsilon'/\epsilon<3$). The corresponding $fZ^2$
at point A and B are $1/(2\sqrt{2}\pi\Gamma^2(l_{\rm B}/a)^2)$ and
$(1-\Gamma)/(\sqrt{2}\pi\Gamma^2(l_{\rm B}/a)^2)$, respectively.
For strongly charged polyelectrolytes such that
$fZ^2>(1-\Gamma)/(\sqrt{2}\pi\Gamma^2(l_{\rm B}/a)^2)$, no
adsorbed state exists.}
 \label{phase1.eps}
%\vspace{20pt}
\end{figure}
%%%%%%%%%%%%%%%%%%%%%%%%%%%%%%%%%%%%%%%%%%%%%%%%%%%%%%%%%%%%%%%%%%%%%%%%%%%%

%%%%%%%%%%%%%%%%%%%%%%%%%%%%%%%%%%%%%%%%%%%%%%%%%%%%%%%%%%%%%%%%%%%%%%%%%%%%%%%%%

In many-chain adsorption problem, we focus on the regime $r_{\rm
D, 0} \gg R$. Expand Eq.(\ref{minimize}) in $R/r_{\rm s, 0}$, we
get the surface polyelectrolyte density
\begin{equation}
n \simeq R^{-2} \simeq \frac{\sigma_{\rm c}-\sigma}{2(1-\Gamma)Z
e} + O(\frac{R}{r_{\rm s, 0}})
\end{equation}
As $\sigma\rightarrow\sigma_{\rm c}^-$, the polyelectrolytes are
going to de-sorb, its surface density decreases to zero. During
the de-sorption, the average distance between polyelectrolytes can
be arbitrary large so that the de-sorption happens as a
single-chain de-sorption process. By comparing two length scales
$R$ and $r_{\rm s, 0}$, the problem is classified into two
regimes. When $r_{\rm s, 0} \gtrsim R$, it corresponds to
many-chain adsorption,
\begin{equation} \label{many}
\sigma \lesssim \sigma_{\rm c}-\frac{2(1-\Gamma)Z e}{r_{\rm s,
0}^2} \simeq \sigma_{\rm c}
\end{equation}
Otherwise, when $r_{\rm s, 0} \lesssim R$, it becomes single chain
adsorption,
\begin{equation} \label{single}
\sigma \gtrsim \sigma_{\rm c}-\frac{2(1-\Gamma)Z e}{r_{\rm s,
0}^2}\simeq \sigma_{\rm c}
\end{equation}
When $\sigma$ increases, inter-chain distance $R$ becomes larger,
and then the system switches from many-chain to single-chain
adsorption. $\sigma_{\rm c}$ is nothing but a surface charge
density indicating the crossover. The threshold surface charge
density is the same as the single chain case. The adsorption
transition is still first-order.

After the strongly correlated polyelectrolytes settled down on the
surface, the effective surface charge felt by monovalent
counterions varies. The new Gouy-Chapman length
\begin{eqnarray} \label{gc}
\lambda &\simeq& \frac{e}{2\pi (1-\Gamma)l_{\rm B}(\sigma + Z e
n)} \nonumber \\ &=& \frac{e}{\pi l_{\rm B}((1-2\Gamma)\sigma +
\sigma_{\rm c})}
\end{eqnarray}
Naive thinking from the above expression, if the surface Debye
screening length $r_{\rm s, 0}$ is fixed, the Gouy-Chapman length
decreases (increases) with increasing surface charge density
$\sigma$ for $\Gamma<1/2$ ($\Gamma>1/2$). However, the naive
picture does not necessarily hold since the counterions may
condense on the surface, and the surface Debye screening length
may be greatly reduced. Applying the Poisson-Boltzmann theory (see
details in Appendix), the problem is again divided into two cases
according to the magnitude of the new Gouy-Chapman (GC) length.

%%%%%%%%%%%%%%%%%%%%%%%%%%%%%%%%%%%%%%%%%%%%%%%%%%%%%%%%%%%%%%%%%%%%%%%%%%%%%%

%%%%%%%%%%%%%%%%%%%%%%%%%%%%%%%%%%%%%%%%%%%%%%%%%%%%%%%%%%%%%%%%%%%%%%%%%%%%
\vspace{20pt}
\begin{figure}[tbh]
\begin{center}
\includegraphics[width=3in]{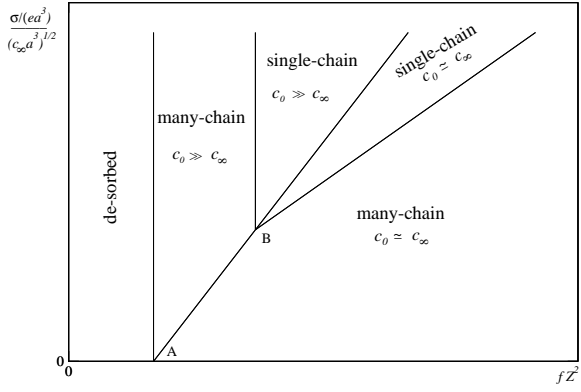}
\end{center}
\vspace{-5pt} \caption{Phase diagram of the adsorbed layer for
$\Gamma>1/2$ (or $\epsilon'/\epsilon>3$). The corresponding $fZ^2$
at point A and B are $(1-\Gamma)/(\sqrt{2}\pi\Gamma^2(l_{\rm
B}/a)^2)$ and $1/(2\sqrt{2}\pi\Gamma^2(l_{\rm B}/a)^2)$,
respectively. For weakly charged polyelectrolytes such that
$fZ^2<(1-\Gamma)/(\sqrt{2}\pi\Gamma^2(l_{\rm B}/a)^2)$, no
adsorbed state exists.}
 \label{phase2.eps}
%\vspace{-5pt}
\end{figure}
%%%%%%%%%%%%%%%%%%%%%%%%%%%%%%%%%%%%%%%%%%%%%%%%%%%%%%%%%%%%%%%%%%%%%%%%%%%%%%%%%

%%%%%%%%%%%%%%%%%%%%%%%%%%%%%%%%%%%%%%%%%%%%%%%%%%%%%%%%%%%%%%%%%%%%%%%%%%%%%%%%%

Case 1). Small GC length; that is, $\lambda<(2\pi l_{\rm
B}c_\infty)^{-1/2}$, where $c_\infty$ is the bulk counterion
density. The surface counterion density
\begin{equation}
c_0 \simeq \frac{1}{2\pi l_{\rm B}\lambda^2} \simeq \frac{\pi
l_{\rm B}}{2e^2}((1-2\Gamma)\sigma + \sigma_{\rm c})^2
\end{equation}
which is much larger than the bulk value $c_\infty$. Surface Debye
screening length
\begin{equation} \label{r0}
r_{\rm s, 0}\simeq (4\pi l_{\rm B}c_0)^{-1/2} \simeq
\frac{e}{\sqrt{2}\pi l_{\rm B}((1-2\Gamma)\sigma+\sigma_{\rm c})}
\end{equation}
Notice that the polyelectrolytes contribution to the ionic
strength is ignored because the polyelectrolytes form a
Wigner-crystal like structure with its neighbors. It cannot
provide significant contribution to mobile screening.

Eliminate $\sigma_{\rm c}$ from Eqs.(\ref{sigmac}) and (\ref{r0}),
\begin{equation} \label{r0b}
r_{\rm s, 0} \simeq
\frac{1}{(1-2\Gamma)(\sigma/e)}\left(\frac{1}{\sqrt{2}\pi l_{\rm
B}}-\frac{\Gamma^2l_{\rm B}f Z^2}{(1-\Gamma)a^2}\right)
\end{equation}
$r_{\rm s, 0}$ is a decreasing function with $fZ^2$ for fixed
$\Gamma$. Fixing $fZ^2$, when $\Gamma$ is increased up to 1/2, the
prefactor in Eq.(\ref{r0b}) turns to be negative. For high enough
$\Gamma$, the bracket turns sign. Suppose we estimate the
adsorption threshold for $fZ^2$ as $r_{\rm s, 0}$ approaching to
zero, then the adsorption happens when
\begin{equation}
fZ^2\lesssim\frac{1-\Gamma}{\Gamma^2}\frac{1}{\sqrt{2}\pi(l_{\rm
B}/a )^2}
\end{equation}
for $\Gamma<1/2$, and it happens when
\begin{equation}
fZ^2\gtrsim\frac{1-\Gamma}{\Gamma^2}\frac{1}{\sqrt{2}\pi(l_{\rm
B}/a )^2}
\end{equation}
for $\Gamma>1/2$.

For high enough $f Z^2$, when $r_{\rm s, 0}$ meets $R$, many-chain
adsorption becomes a single-chain problem. By Eq.(\ref{single})
and (\ref{r0b}), we have
\begin{equation}
f Z^2 \lesssim\frac{1}{2\sqrt{2}\pi\Gamma^2(l_{\rm B }/a)^2}
\end{equation}
for many-chain adsorption. Finally, for self-consistency, we need
to check the condition $\lambda<(2\pi l_{\rm B}c_\infty)^{-1/2}$,
by Eq.(\ref{gc}), (\ref{r0}), and (\ref{r0b}), where it gives
\begin{eqnarray}
\frac{\sigma/(ea^{-2})}{(c_\infty a^3)^{1/2}}>
\frac{2\sqrt{\pi}\Gamma^2(l_{\rm B
}/a)^{3/2}}{(1-2\Gamma)(1-\Gamma)}\left(
\frac{1-\Gamma}{\sqrt{2}\pi\Gamma^2(l_{\rm B }/a)^2}-fZ^2\right)
\nonumber \\
\end{eqnarray}
The regime corresponds to high surface charge density or low bulk
counterion density.

Case 2). Large GC length; that is, $\lambda>(2\pi l_{\rm
B}c_\infty)^{-1/2}$. According to Poisson-Boltzmann theory, the
surface counterion density is almost the same as the bulk one
under this condition. The surface Debye screening length
\begin{equation}
r_{\rm s, 0} \simeq (4\pi l_{\rm B}c_\infty)^{-1/2}
\end{equation}
which is independent of $\sigma$. For many-chain adsorption,
Eq.(\ref{sigmac}) and  (\ref{many}) gives
\begin{equation}
f Z^2 \gtrsim \frac{1-\Gamma}{2\sqrt{\pi}\Gamma^2(l_{\rm B}/a
)^{3/2}}\frac{\sigma/(ea^{-2})}{(c_\infty a^3)^{1/2}}
\end{equation}
Self-consistent condition $\lambda>(2\pi l_{\rm
B}c_\infty)^{-1/2}$ requires
\begin{eqnarray}
\frac{\sigma/(ea^{-2})}{(c_\infty a^3)^{1/2}}<
\frac{2\sqrt{\pi}\Gamma^2(l_{\rm B
}/a)^{3/2}}{(1-2\Gamma)(1-\Gamma)}\left(
\frac{1-\Gamma}{\sqrt{2}\pi\Gamma^2(l_{\rm B }/a)^2}-fZ^2\right)
\nonumber \\
\end{eqnarray}

In summary, according to the dielectric ratio $\epsilon'/\epsilon$
(or the coupling strength $\Gamma$), the phase diagrams of the
adsorbed layer are divided into two classes as shown in
Fig.\ref{phase1.eps} and \ref{phase2.eps}. When the surface charge
density is low (or the bulk counterion density is high), the
surface and the bulk counterion density are almost the same. Once
the surface charge density is high enough ($\lambda<(2\pi l_{\rm
B}c_\infty)^{-1/2}$), counterions condense on the surface. In this
regime, polyelectrolytes of lower valency
($fZ^2<1/(2\sqrt{2}\pi\Gamma^2(l_{\rm B}/a)^2)$) form a correlated
many-chain state. As their valency is high enough, the state turns
out to be single-chain because of stronger repulsion between
neighboring chains.

The qualitative differences between the phase diagram of
$\Gamma<1/2$ (or $\epsilon'/\epsilon<3$) and $\Gamma>1/2$ (or
$\epsilon'/\epsilon>3$) is the following. Starting from the
many-chain state with $c_0\simeq c_\infty$, it is expected the
system transits to the single-chain state when the surface charge
density increases. For $\Gamma<1/2$, it happens for weakly charged
polyelectrolytes ($fZ^2<1/(2\sqrt{2}\pi\Gamma^2(l_{\rm B}/a)^2)$);
while for $\Gamma>1/2$, strongly charged polyelectrolytes are
needed. Another difference is their de-sorbed states. For
$\Gamma<1/2$, no adsorbed state exists for strongly charged
polyelectrolytes ($fZ^2>(1-\Gamma)/(\sqrt{2}\pi\Gamma^2(l_{\rm
B}/a)^2)$); while weakly charged polyelectrolytes are all
de-sorbed for $\Gamma>1/2$.

We emphasize that our treatment of multi-valent polyelectrolytes
interaction within the Gouy-Chapman layer is preliminary, in which
it is still subtle \cite{sens,netz2}. The qualitative prediction
of the phase diagram should not be altered and could be tested
experimentally.

The work is supported by National Science Council of Republic of
China under Grant No. NSC91-2816-M-008-0009-6 (CHC) and
NSC92-2112-M-008-051 (PYL).

\section{Appendix: Poisson-Boltzmann Theory for high-$Z$}

Suppose the surface charge density at the surface $\sigma > 0$.
The polyion distribution of valency $Z$ carrying charge $Ze$ is
$\rho(z)$, and its counterion distribution of monovalency is
$c(z)$. Neglecting the ion correlation in Poisson-Boltzmann
theory, we have
\begin{eqnarray}
\rho(z) &=& \rho_{\infty}\exp(-\beta Ze \psi(z))  \\
c(z) &=& c_{\infty}\exp(\beta e \psi(z))
\end{eqnarray}
$c_{\infty}=Z \rho_{\infty}$ neutralizes the whole system. The
Poisson-Boltzmann equation becomes
\begin{equation} \label{pb1}
\frac{d^2\psi}{dz^2}=\frac{4\pi
ec_{\infty}}{\epsilon}\left(\exp(\beta e\psi)-\exp(-Z \beta e
\psi)\right)
\end{equation}
In general, only a few solution can be expressed in a closed form
(e.g. $Z=1, 2)$ \cite{grahame}. For $Z\rightarrow\infty$, the
above equation is reduced to
\begin{equation}
\frac{d^2\psi}{dz^2}=\frac{4\pi ec_{\infty}}{\epsilon}\exp(\beta
e\psi)
\end{equation}
with the boundary condition
%\begin{equation}
$\left.\psi'\right|_{z=0}=-4\pi\sigma/\epsilon$.
%\end{equation}
Near the surface $z=0$, it gives the Gouy-Chapman form
\begin{equation} \label{gouy-chapman}
c(z)=\frac{A}{(z+\lambda)^2}
\end{equation}
where $\lambda=e/(2\pi l_{\rm B}\sigma)$ and $A$ is an
undetermined constant. In the usual Gouy-Chapman solution,
$A=1/(2\pi l_{\rm B})$ because we restrict
$c(z\rightarrow\infty)=0$. In our case
$c(z\rightarrow\infty)=c_{\infty}>0$ and hence the Gouy-Chapman
form is only valid near the surface.

In term of the variable $c(z)$, Eq.(\ref{pb1}) can be written as
\begin{equation} \label{pb2}
\frac{1}{c}\frac{d^2c}{dz^2}-\frac{1}{c^2}\left(\frac{dc}{dz}\right)^2
= 4\pi l_{\rm B} \left( c -
c_\infty\left(\frac{c_\infty}{c}\right)^Z\right)
\end{equation}
In order to consider the solution at the limit of
$Z\rightarrow\infty$, we need self-consistency condition
$c_\infty/c_0\leq 1$. Substitute Eq.(\ref{gouy-chapman}) into
Eq.(\ref{pb2}) and restrict $z\simeq 0$, one gets $A \simeq
1/(2\pi l_{\rm B})$ if $\lambda < (2\pi l_{\rm
B}c_\infty)^{-1/2}$, and $A \simeq c_\infty\lambda^2$ otherwise.

The above result are derived from the high-$Z$ limit. Now we are
going to look at the next order contribution from finite $Z^{-1}$.
Write the electric potential into $\psi = \psi_0 + \psi_1$ such
that $\psi_0$ is the solution shown above, and $\psi_1$ is the
next order correction subject to the following boundary condition
%\begin{equation}
$\left.\psi_1\right|_{z=0} = \left.\psi_1'\right|_{z=0} = 0$.
%\end{equation}
Expand in $\psi_1$ and around $z=0$,
\begin{eqnarray}
\frac{d^2\psi_1}{dz^2}&=& -\frac{4\pi e
c_\infty}{\epsilon}\left({\rm e}^{-Z\beta e\psi_0} -({\rm
e}^{\beta e\psi_0} +Z {\rm e }^{-Z\beta e\psi_0})\beta e\psi_1
\right)
\nonumber \\
&\simeq& -\frac{4\pi e c_\infty}{\epsilon}\left(\frac{A}{c_\infty
\lambda^2}\right)^Z
\end{eqnarray}
if we keep only the leading order. Solving for $\psi_1$ gives
\begin{eqnarray}
c(z) &=& c_\infty \exp(\beta e(\psi_0+\psi_1)) \nonumber
\\
&\simeq& \frac{A}{(z+\lambda)^2}\left(1-2\pi l_{\rm B}c_\infty
\left(\frac{c_\infty\lambda^2}{A}\right)^Z z^2\right)
\end{eqnarray}

In summary, for $Z\gg 1$, when $\lambda < (2\pi l_{\rm
B}c_\infty)^{-1/2}$,
\begin{equation}
c(z)\simeq \frac{1}{2\pi l_{\rm B}(z+\lambda)^2}\left(1 - (2\pi
l_{\rm B}c_\infty)^{Z+1} \lambda^{2Z} z^2 \right)
\end{equation}
at $z\ll\lambda$. Hence $c_0 \simeq 1/(2\pi l_{\rm B}\lambda^2)
\gg c_\infty$ implying that counterions condense on the surface.
On the other hand, when $\lambda>(2\pi l_{\rm B}c_\infty)^{-1/2}$,
\begin{equation}
c(z)\simeq\frac{c_\infty \lambda^2}{(z+\lambda)^2}\left(1-2\pi
l_{\rm B}c_\infty z^2\right)
\end{equation}
at $z \ll (2\pi l_{\rm B}c_\infty)^{-1/2}$. Then $c_0 \simeq
c_\infty$.

\newpage

\newpage

\end{document}